\documentclass[useAMS,usenatbib]{mn2e}
\usepackage{graphicx}

\usepackage[usenames,dvipsnames]{xcolor}
\usepackage{url}
\usepackage{amssymb}
\usepackage{lscape}

\newcommand{\kms}{~km s$^{-1}$~}
\newcommand{\WR}{WR~30a~}
\newcommand{\WRE}{WR~30a}
\newcommand{\dotM}{~M$_{\odot}$~yr$^{-1}$~}
\newcommand{\XMM}{{\it XMM-Newton~}}
\newcommand{\XMME}{{\it XMM-Newton}}
\newcommand{\Chandra}{{\it Chandra~}}

\newcommand{\Rosat}{{\it ROSAT~}}
\newcommand{\RosatE}{{\it ROSAT}}
\newcommand{\apj}{ApJ}
\newcommand{\xspec}{{\sc xspec~}}
\newcommand{\xspecE}{{\sc xspec}}

\def\kms{\mbox{~km\,s$^{-1}$\/}}

\def\arcsec{\hbox{$^{\prime\prime}$}}
\def\utw{\smash{\rlap{\lower5pt\hbox{$\sim$}}}}
\def\udtw{\smash{\rlap{\lower6pt\hbox{$\approx$}}}}

\title[X-rays from \WR]
{X-rays from the oxygen-type Wolf-Rayet binary  WR30a}
\author[S.A.Zhekov and S.L.Skinner]{Svetozar A. Zhekov$^1$\thanks{
E-mail: szhekov@astro.bas.bg; stephen.skinner@colorado.edu.} 
and Stephen L. Skinner$^2$\\
$^1$Institute of Astronomy and National Astronomical Observatory, 
72 Tsarigradsko Chaussee Blvd., Sofia 1784, Bulgaria\\
$^2$CASA, University of Colorado, Boulder, CO
80309-0389, USA
}

\date{}

\pagerange{\pageref{firstpage}--\pageref{lastpage}} \pubyear{2002}

\begin{document}

\maketitle

\label{firstpage}

\begin{abstract}
We present an analysis of  \XMM X-ray data of \WRE ~(WO$+$O), a close 
massive binary that harbours an oxygen-rich Wolf-Rayet star.
Its spectrum is characterized by the presence of two well-separated 
broad peaks, or `bumps', one peaking at energies between 1 and 2 keV 
and the other between 5 and 7 keV. A two-component model is required 
to match the observed spectrum. The higher energy spectral peak  is
considerably more absorbed and dominates the X-ray luminosity.
For the currently accepted distance of 7.77 kpc, the X-ray
luminosity of \WRE ~is $L_X > 10^{34}$ erg s$^{-1}$, making it
one of the 
most X-ray luminous WR$+$O binary amongst those in the Galaxy
with orbital periods less than $\sim20$ d.
The X-ray spectrum  can be acceptably fitted using either thermal or
nonthermal models, so the X-ray production mechanism is yet unclear.

\end{abstract}


\section{Introduction}
\label{sec:intro}
Wolf-Rayet (WR) stars are  massive stars that have powerful winds 
and are losing mass at high rates 
($\dot{M}$ $\sim$ 10$^{-5}$ \dotM;
V$_{wind} = 1000-5000$\kms). 
Based on their optical spectra, they are divided into three subtypes:
nitrogen-rich (WN),  carbon-rich (WC), and oxygen-rich (WO).
Their progenitors are massive stars with initial masses 
$>$25 M$_{\odot}$ and most WR stars are thought to  end their lives as 
supernovae (see \citealt{crowther_07} for a  review on the physical 
properties of WRs).
The observed binary fraction in Galactic Wolf-Rayet stars  is
relatively high with about 40\% being members of WR$+$O systems
\citep{vdh_01}.

WR stars were discovered to be X-ray sources by the
{\it Einstein Observatory} \citep{seward_79}.
The first systematic survey of WRs showed that  WR$+$O binaries
are the brightest X-ray sources amongst them \citep{po_87}.
Their enhanced emission likely originates from the interaction region 
of the winds of the massive binary components
(\citealt{pri_us_76}; \citealt{cherep_76}).

After the launch of the modern X-ray observatories \Chandra and
\XMME, the number of WR stars with good quality X-ray spectra has 
increased considerably. Thus, some similarities and differences in 
the characteristics of their X-ray emission are now  established.

Most WR stars are classified as either WC or WN subtypes and analysis 
of undispersed spectra of presumably single objects have revealed  
clear differences between WC and WN stars. All observations of  
putatively single WC stars so far have resulted in non-detections
(\citealt{os_03}; \citealt{sk_06}). In contrast, WN stars without 
known companions are detected in X-rays and their spectra typically 
reveal an admixture of cool (kT $< 1$~keV) and hot (kT $> 2$~keV) 
plasma (\citealt{sk_10}, 2012).
WO stars are much rarer. The only pointed X-ray observations of 
a presumably single Galactic  WO star to date, WR 142, resulted
in a detection (\citealt{os_09}; \citealt{sokal_10}).
Interestingly, the \Chandra detection of WR 142 showed an extremely hard
and heavily-absorbed X-ray spectrum which could be fitted using
either thermal or nonthermal models \citep{sokal_10}.

On the other hand, the X-ray properties of massive WR$+$O binaries 
do not seem to have any strong dependence
on the subtype of the WR component. Analyses of the X-ray 
emission from both close and wide WR$+$O binaries  give support to the 
idea that a substantial fraction of their X-ray emission arises in
colliding stellar wind (CSW) shocks, resulting from the
interaction of the massive winds of the binary components
(e.g., \citealt{sk_01}; \citealt{raasen_03}; \citealt{schild_04}; 
\citealt{po_05}; \citealt{zhp_10a}, b; \citealt{zhgsk_11}, 2014;
\citealt{zh_12}).
But  no WR$+$O binary with an oxygen-rich WR component
has heretofore been  detected in X-rays.

\begin{figure*}
\begin{center}
\centering\includegraphics[width=1.5in,height=2in,angle=-90] {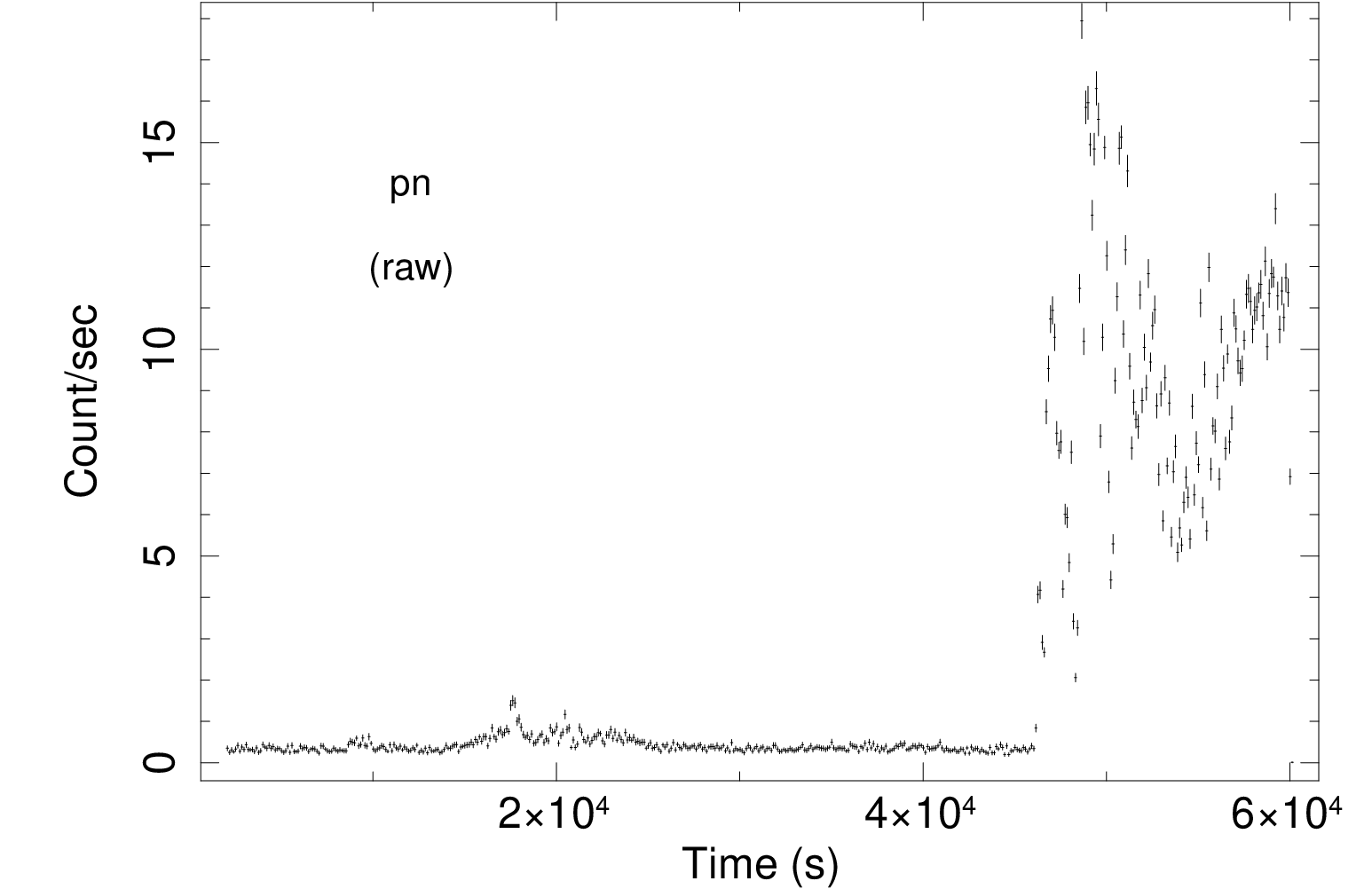}
\centering\includegraphics[width=1.5in,height=2in,angle=-90] {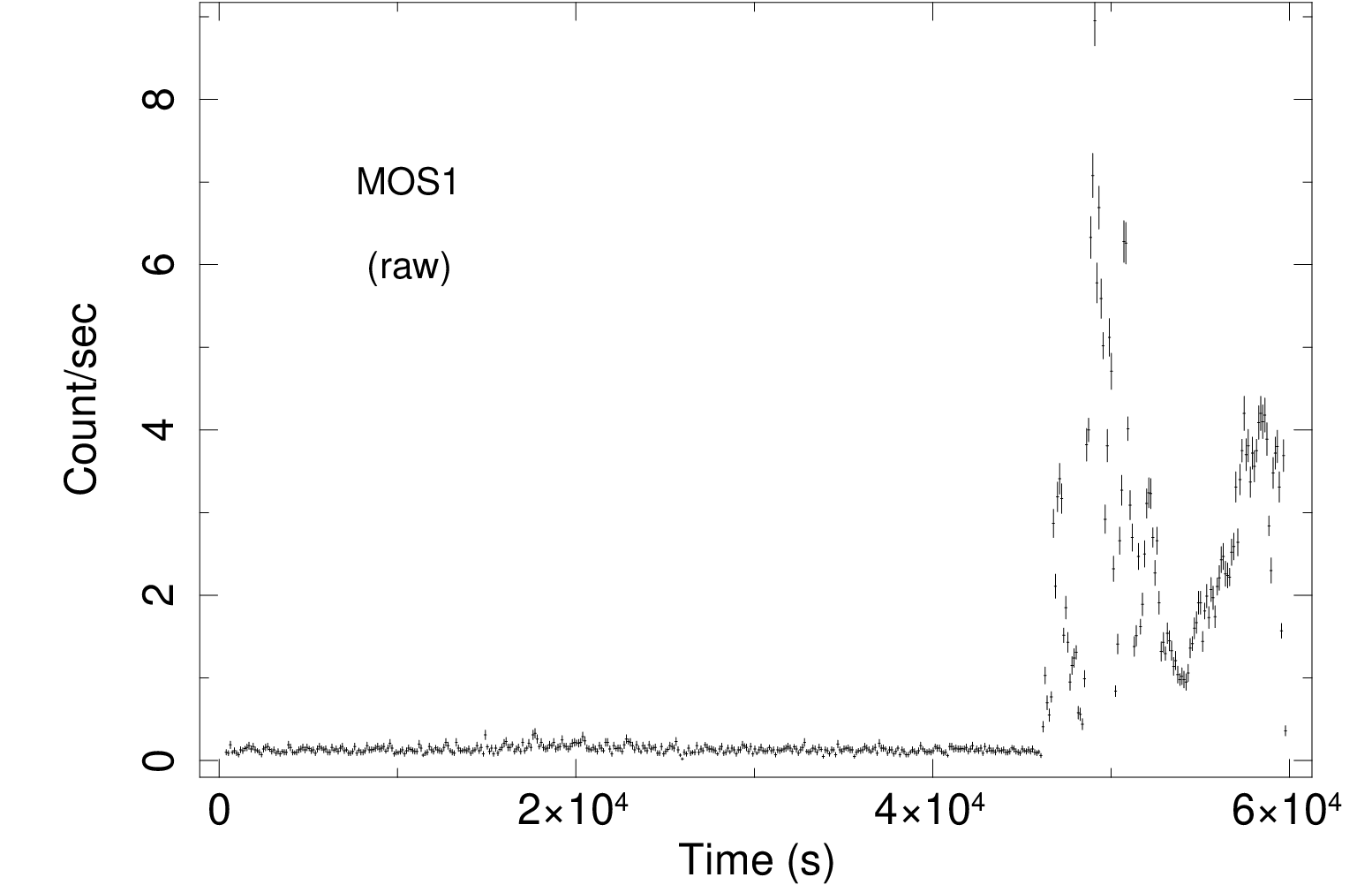}
\centering\includegraphics[width=1.5in,height=2in,angle=-90] {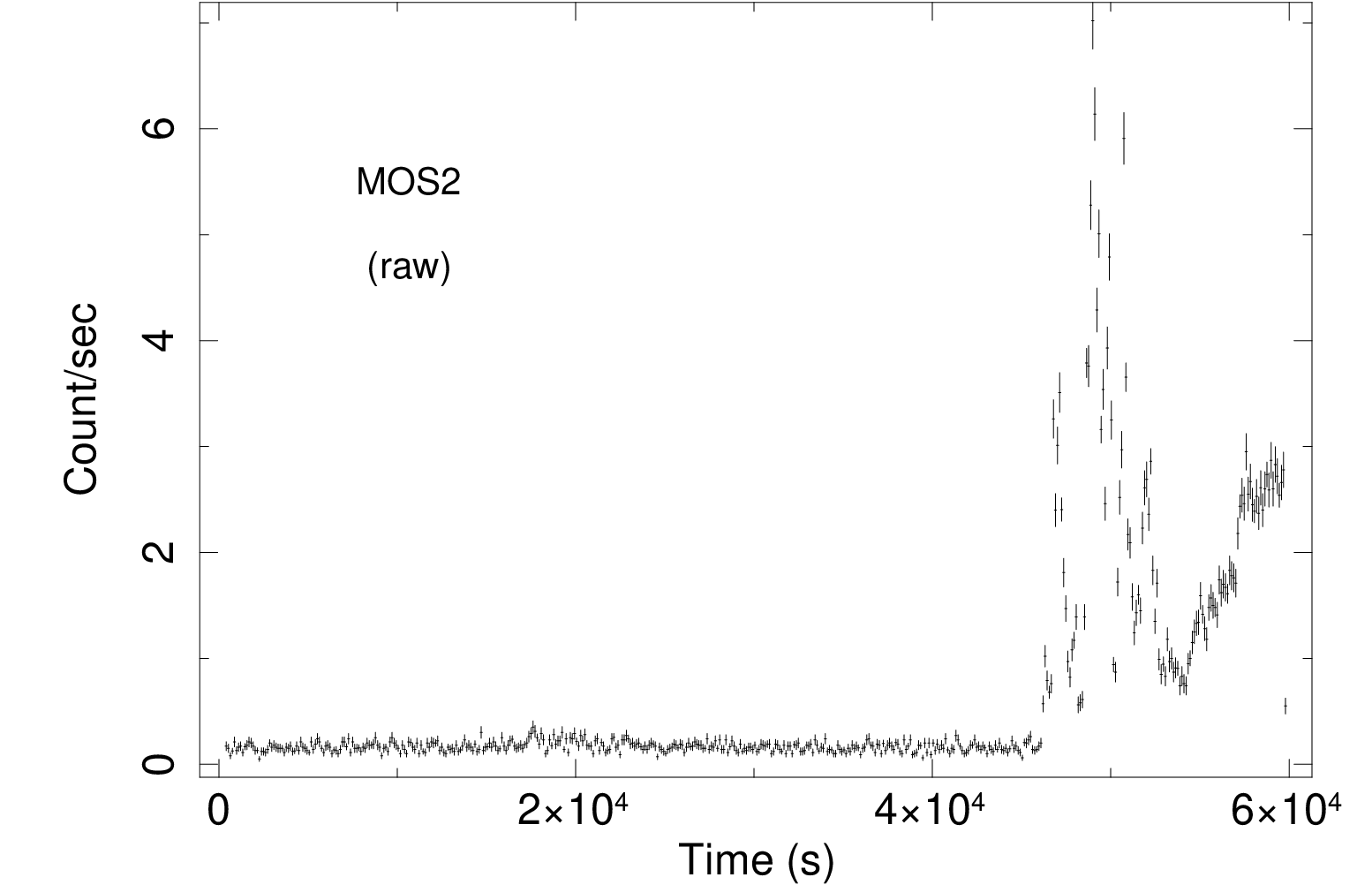}
\centering\includegraphics[width=1.5in,height=2in,angle=-90] {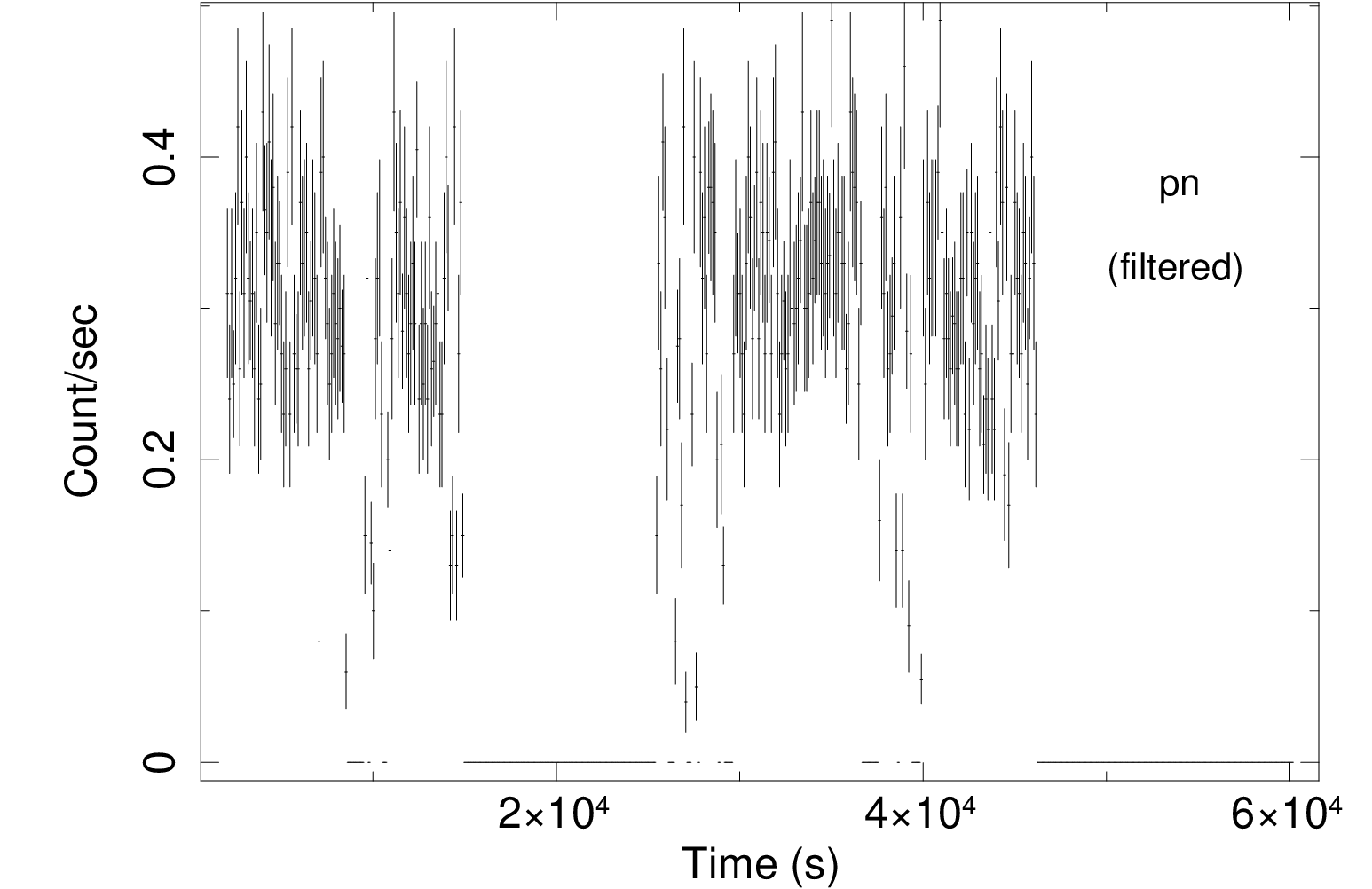}
\centering\includegraphics[width=1.5in,height=2in,angle=-90] {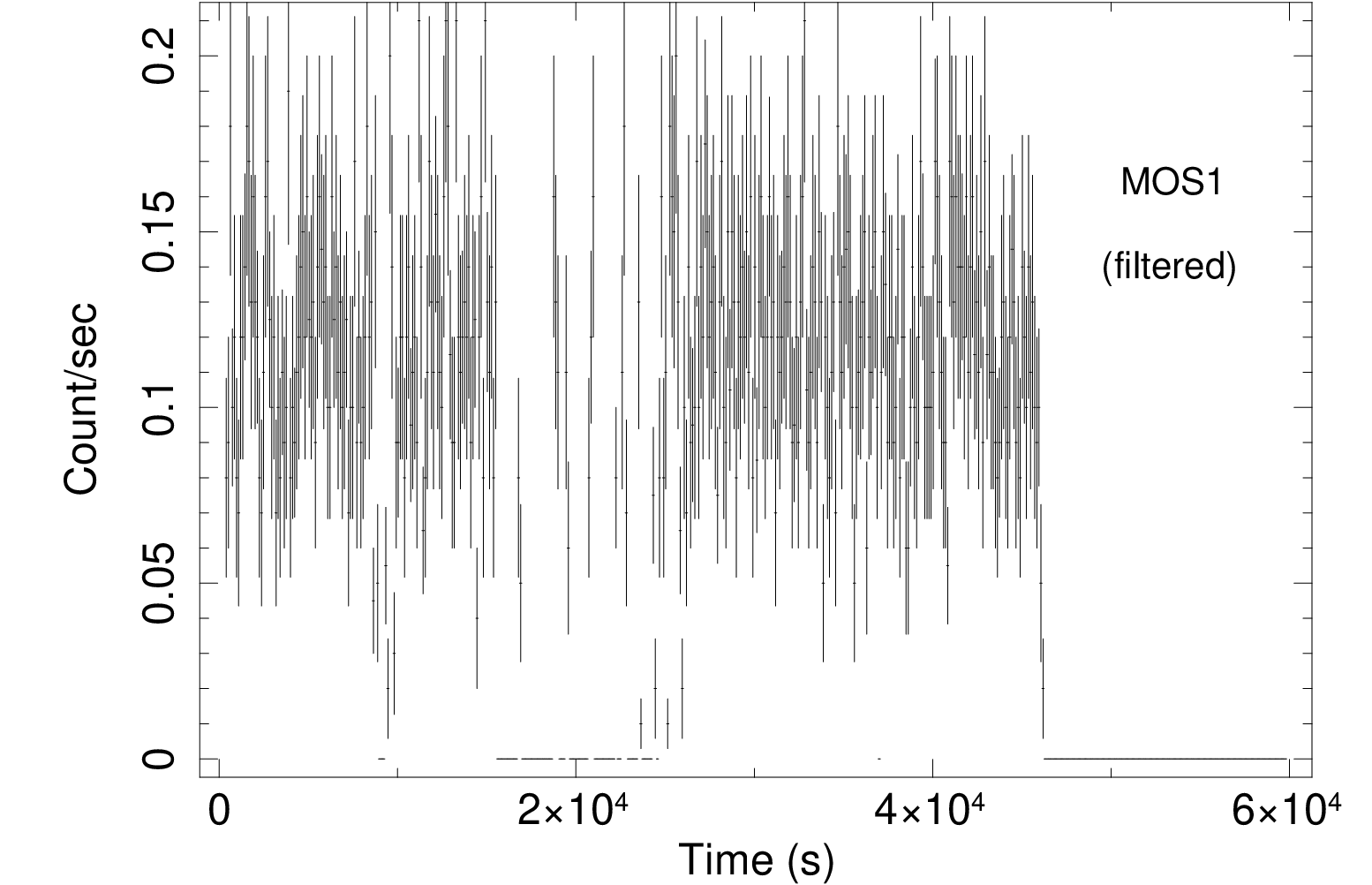}
\centering\includegraphics[width=1.5in,height=2in,angle=-90] {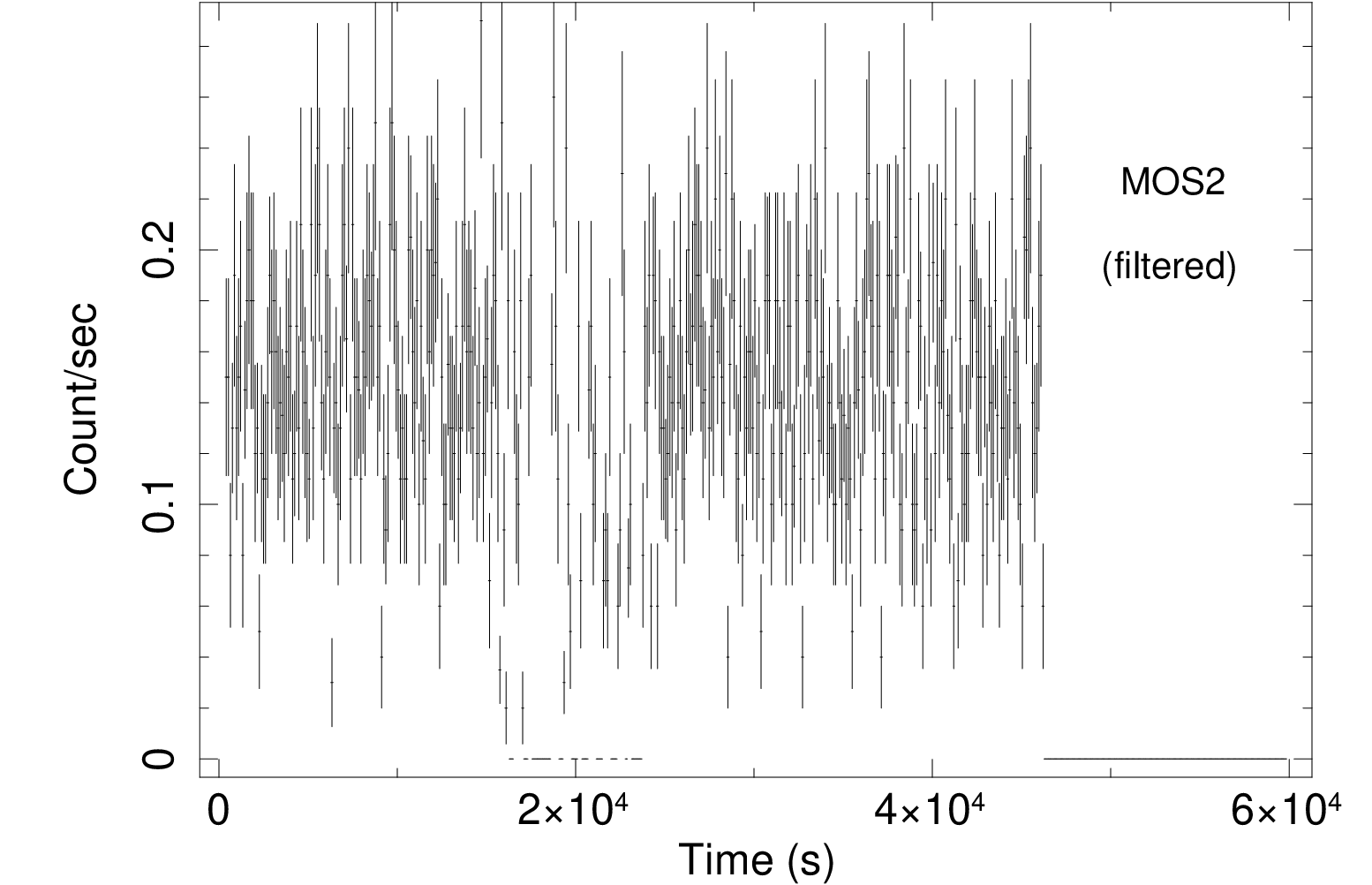}
\centering\includegraphics[width=2in,height=1.734in] {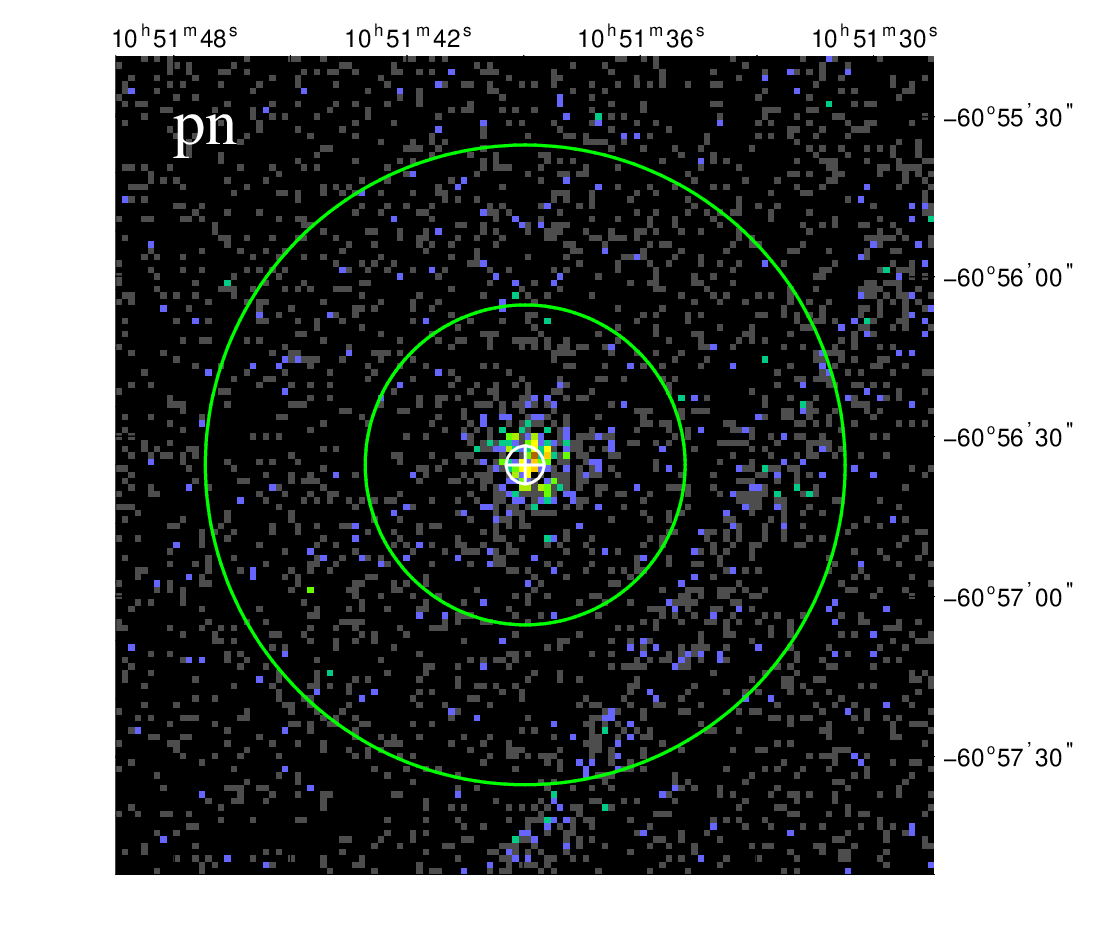}
\centering\includegraphics[width=2in,height=1.734in] {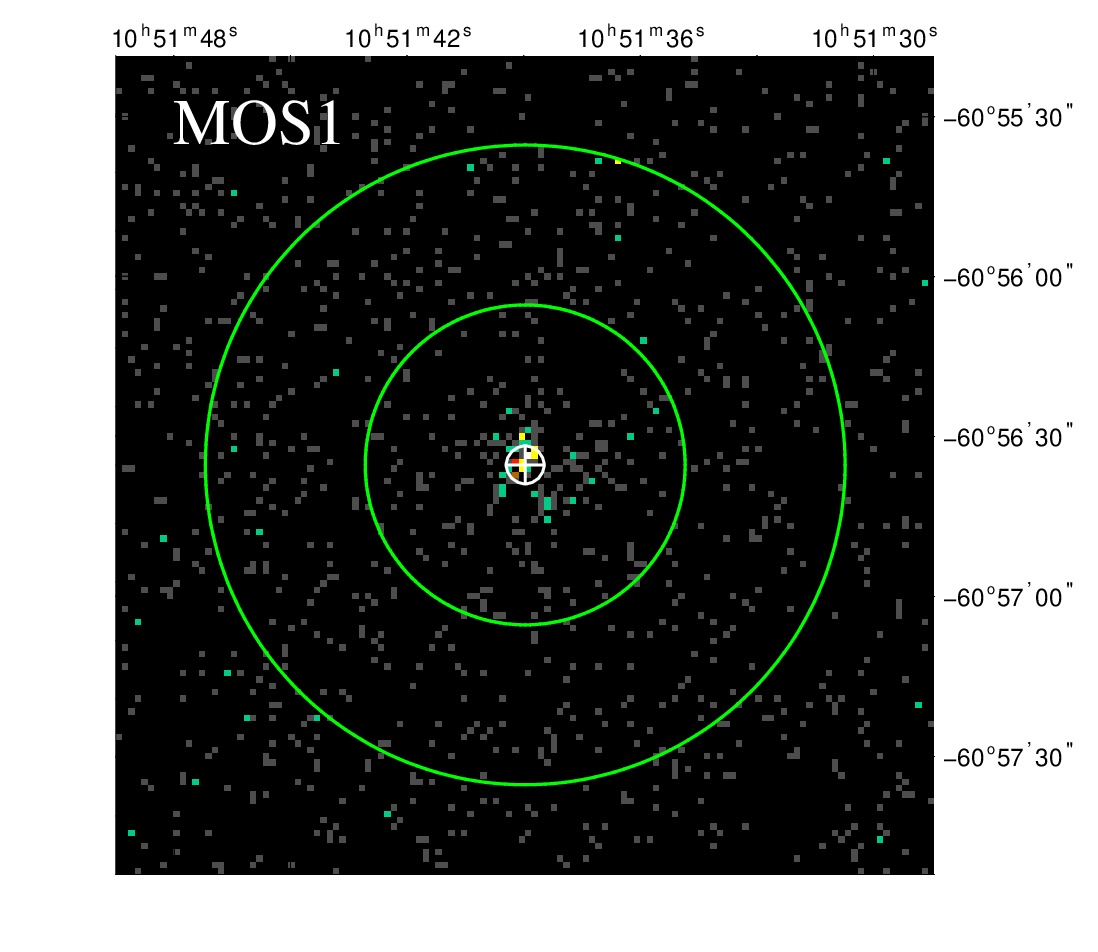}
\centering\includegraphics[width=2in,height=1.734in] {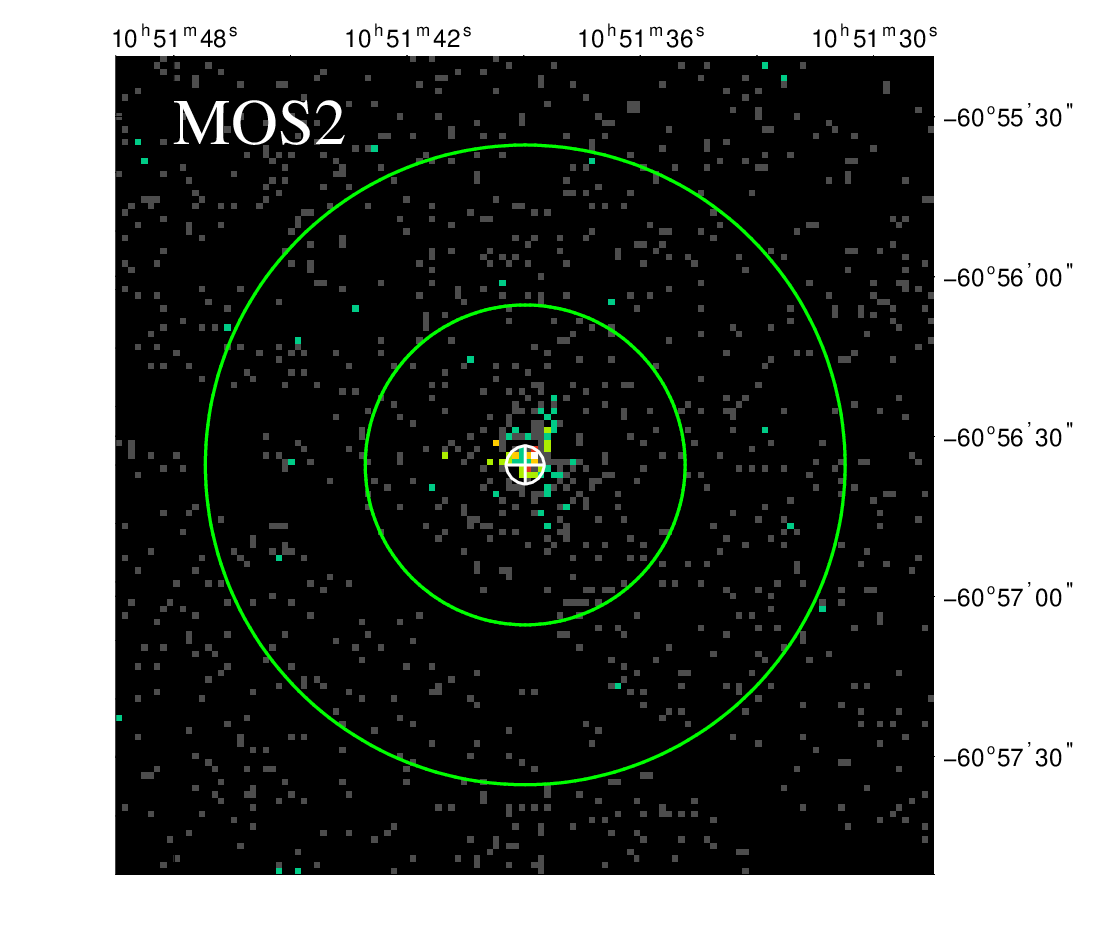}
\end{center}
\caption{
{\it Top row.} The total-field high-energy ($>$10 keV) EPIC light 
curves of \WRE.
{\it Middle row.} The total field high-energy EPIC light curves of \WR 
after being  {\it filtered} to remove  high X-ray background.
{\it Bottom row.} The {\it filtered} EPIC images of \WR in the 
0.2-10 keV energy band. The 
source spectrum was extracted from within the central circle 
(radius of 30\arcsec). The  background spectrum was extracted from 
the surrounding annulus (outer radius of 60\arcsec). The circled 
plus sign gives the optical position of \WR (SIMBAD).
}
\label{fig:image}
\end{figure*}

In this paper, we report the X-ray detection of the close WO$+$O 
binary system \WRE.
The paper is organized as follows.
We summarize  information on  \WR
in Section~\ref{sec:wrstar}. In Section~\ref{sec:observations}, we
review the X-ray observations and data. In Section~\ref{sec:results}, 
we present results from analysis of the X-ray properties of \WRE. In
Section~\ref{sec:discuss}, we discuss our results, and we present our
conclusions in Section~\ref{sec:conclusions}.

\section{The Wolf-Rayet Star \WR}
\label{sec:wrstar}
\WR (V574 Car; [MS70])
is one of  four known oxygen-rich WR stars in the Galaxy
(\citealt{vdh_01}, 2006; also  
Galactic Wolf-Rayet Catalogue\footnote{\url{http://pacrowther.staff.shef.ac.uk/WRcat/index.php}}).
It is a WO4$+$O5-5.5 binary system at a distance of 7.77 kpc 
\citep{vdh_01}. Detailed photometric and spectroscopic studies proposed 
an orbital period of 4.619 d and classified the O5 component as either
a main-sequence or, more likely, a giant star \citep{gosset_01}.
The optical extinction toward \WR is A$_v = 4.26$~mag
(\citealt{vdh_01}; A$_v = 1.11$ A$_{\mbox{V}}$) 
implying a foreground column density of
N$_H = (6.33-8.52)\times10^{21}$~cm$^{-2}$.
The range corresponds to the conversion that is used:
N$_H = (1.6-1.7)\times10^{21}$A$_{\mbox{V}}$~cm$^{-2}$
(\citealt{vuong_03}, \citealt{getman_05});
and 
N$_H = 2.22\times10^{21}$A$_{\mbox{V}}$~cm$^{-2}$
\citep{go_75}.
Similarly to other WO stars, \WR has a very fast stellar wind: 
V$_{wind} = 4500$\kms \citep{kings_95}.

No radio data on \WR were  found  in the literature.
\citet{po_95} reported an X-ray non-detection in the {\it \Rosat}
PSPC Survey on Wolf-Rayet stars, likely due to the very 
short exposure time (260 s).

\section{Observations and data reduction}
\label{sec:observations}
We searched the \Chandra and \XMM archives for data on \WR and 
found one X-ray data set. 
\WR was observed with \XMM on 2010 January 25 (Observation ID
0606330101) with a nominal exposure of $\sim 60$ ks.
Our study is based on the data from the European Photon Imaging Camera 
(EPIC), which has  one pn and two nearly-identical MOS detectors\footnote{See \S~3.3 in the 
\XMM Users Handbook, Issue 2.12, 2014 (ESA: XMM-Newton SOC)
\url{http://xmm.esac.esa.int/external/xmm_user_support/documentation/uhb}
}.
We used the \XMM 
{\sc sas}\footnote{Science Analysis Software, 
\url{http://xmm.esac.esa.int/sas}} 14.0.0 data analysis software for 
data reduction, following the procedures summarized  below.

First, the {\sc sas} pipeline processing scripts {\em emproc} and {\em epproc}
were executed to incorporate the most recent calibration files (as of
2015 January 27). The data were checked for high X-ray background 
following the instructions in the {\sc sas} 
documentation\footnote{see \S~4.4.4 in Users Guide to the XMM-Newton 
Science Analysis System, Issue 11.0, 2014 (ESA: XMM-Newton SOC);
\url{http://xmm.esac.esa.int/external/xmm_user_support/documentation/sas_usg/USG/}}, namely, by
constructing total-field light curves at high energies: E $> 10$ keV
for MOS1,2 and  E in the 10 - 12 keV range for pn.
As seen from Fig.~\ref{fig:image}, a high (and flaring) X-ray 
background was registered in the EPIC exposure, with the  pn data
being most affected.
Nevertheless, we applied a background filtering  procedure that 
allowed us to make use of
some of the pn exposure and a substantial fraction of the MOS exposures.

We made use of the \XMM Extended  Source Analysis 
Software ({\sc xmm-esas}) 
package\footnote{\url{http://xmm.esac.esa.int/sas/current/doc/esas}},
which is now incorporated in {\sc sas}.
We created filtered event files using the {\sc esas} commands
{\em mos-filter} and {\em pn-filter} which minimize the contamination 
of the soft proton flaring in the EPIC data in a robust manner
(for details see the {\sc xmm-esas} 
Cookbook\footnote{\url{ftp://xmm.esac.esa.int/pub/xmm-esas/xmm-esas.pdf}}).
As a check, we constructed the total-field filtered light curves at high
energies (Fig.~\ref{fig:image}) which provided confidence that 
the background had been reduced to an accepable level.
 
We then proceeded with the source and background spectral extraction 
from the pn and MOS1,2 filtered event files using the extraction regions  
shown in Fig.~\ref{fig:image}.
The 
{\sc sas} procedures {\em rmfgen} and {\em arfgen} were used to 
generate the corresponding response matrix files and ancillary 
response files for each spectrum. The MOS spectrum used in our 
analysis is the sum of the spectra from the two MOS detectors. 
The extracted spectra (0.2 - 10 keV) of \WR had
$\sim 403$~source counts ($\sim 710$ source$+$background counts) 
in the 24.1-ks pn effective exposure and 
$\sim 350$~source counts ($\sim 584$ source$+$background counts) 
in the 37.2-ks MOS effective exposure (36.0 ks in MOS1 and 38.4 ks in
MOS2).
 
Since the pn exposure was more heavily affected by high X-ray
background, our analysis emphasized  the MOS spectrum  while the pn
spectrum was used to check the consistency of the results.
For the spectral analysis, we used version 12.8.2 of \xspec
\citep{Arnaud96}.

Also, we constructed the pn and MOS1,2 background-subtracted light 
curves  of \WRE. Since the deleted high-background time intervals 
were unevenly-spaced, the  good time intervals (GTIs) in the filtered 
light curves were as well.
There were 27 GTI segments as short as 120 s and as long as 6,540 s
in MOS1; 38 GTI segments as short as 120 s and as long as 4,315
s in MOS; and 26 GTI segments as short as 15 s and as long as 7,080 s
in pn. On a timescale less than 24-37 ks, the X-ray 
light curves were statistically consistent with a constant count rate.  
A constant count rate model gives a goodness-of-fit  $\geq 0.98$
using $\chi^2$ fitting.

\begin{table*}
\caption{Global Spectral Model Results
\label{tab:fits}}
\begin{tabular}{lllllllll}
\hline
\multicolumn{1}{c}{} &
\multicolumn{4}{c} {$\leftarrow -----$ Thermal $ -----\rightarrow$} &
\multicolumn{4}{c} {$\leftarrow -----$ Non-thermal $ -----\rightarrow$} \\
\multicolumn{1}{c}{Parameter} & 
\multicolumn{2}{c}{MOS }  & \multicolumn{2}{c}{MOS$+$pn } &
\multicolumn{2}{c}{MOS }  & \multicolumn{2}{c}{MOS$+$pn } \\
\multicolumn{1}{c}{ } &
\multicolumn{1}{c}{model A}  & \multicolumn{1}{c}{model B} &
\multicolumn{1}{c}{model A}  & \multicolumn{1}{c}{model B} &
\multicolumn{1}{c}{model C}  & \multicolumn{1}{c}{model D} &
\multicolumn{1}{c}{model C}  & \multicolumn{1}{c}{model D} \\
\hline
$\chi^2$/dof  & 
13.1/23 & 19.7/23 & 45.4/58 &  46.4/58  &
13.6/23 & 19.9/23 & 41.1/58 &  45.8/58  \\
N$_{H,1}$ (10$^{22}$ cm$^{-2}$)  &
          0.50$^{+0.20}_{-0.16}$ & 0.852 &
          0.66$^{+0.17}_{-0.16}$ & 0.852 &
          0.78$^{+0.48}_{-0.33}$ & 0.852 &
          1.00$^{+0.33}_{-0.26}$ & 0.852 \\
N$_{H,2}$ (10$^{22}$ cm$^{-2}$)  &
          135.$^{+53.9}_{-34.5}$ &  & 112.$^{+26.9}_{-20.5}$ &  &
          142.$^{+63.3}_{-39.7}$ &  & 118.$^{+0.25}_{-0.16}$ &  \\
N$_{He,1}$ (10$^{22}$ cm$^{-2}$)  &
                                 & 0.00$^{+....}_{-....}$ &
                                 & 0.00$^{+....}_{-....}$ &
                                 & 0.00$^{+0.01}_{-0.00}$ &
                                 & 0.00$^{+0.01}_{-0.00}$ \\
N$_{He,2}$ (10$^{22}$ cm$^{-2}$)  &
                                 & 0.11$^{+0.02}_{-0.02}$ &
                                 & 0.11$^{+0.02}_{-0.02}$ &
                                 & 0.09$^{+0.02}_{-0.02}$ &
                                 & 0.09$^{+0.02}_{-0.01}$ \\
kT (keV) & 
          4.36$^{+4.10}_{-1.33}$ & 2.71$^{+0.88}_{-0.63}$ &
          4.24$^{+3.92}_{-1.22}$ & 2.87$^{+0.81}_{-0.50}$ &
                &   &   &   \\
EM$_1$ ($10^{54}$~cm$^{-3}$) &  
                                0.20$^{+0.04}_{-0.03}$ &
                                0.21$^{+0.03}_{-0.03}$ &
                                0.22$^{+0.03}_{-0.02}$ &
                                0.22$^{+0.03}_{-0.02}$ &
                                 & & & \\
EM$_2$ ($10^{54}$~cm$^{-3}$) &  
                                15.0$^{+21.4}_{-8.33}$ &
                                9.41$^{+10.2}_{-4.20}$ &
                                8.90$^{+6.76}_{-3.83}$ &
                                8.66$^{+5.01}_{-3.23}$ &
                                 & & & \\
$\Gamma_{pow}$ &      &   &   &   &
          2.22$^{+0.68}_{-0.54}$ & 2.41$^{+0.86}_{-0.22}$ &
          2.39$^{+0.48}_{-0.42}$ & 2.70$^{+0.58}_{-0.49}$ \\
norm$_{pow,1}$ (10$^{-4}$)  &      &   &   &   &
          0.25$^{+0.28}_{-0.11}$ & 0.29$^{+0.44}_{-0.06}$ &
          0.32$^{+0.22}_{-0.12}$ & 2.70$^{+0.58}_{-0.49}$ \\
norm$_{pow,2}$ (10$^{-4}$)  &      &   &   &   &
          23.6$^{+76.7}_{-17.2}$ & 7.18$^{+39.2}_{-2.87}$ &
          19.9$^{+34.6}_{-12.0}$ & 14.9$^{+36.8}_{-9.80}$ \\
F$_{X}$ ($10^{-13}$ erg cm$^{-2}$ s$^{-1}$)  &
           \,\,1.87  & \,\,1.73  &
           \,\,1.67  & \,\,1.76  &
           \,\,2.14  & \,\,1.83  &
           \,\,1.91  & \,\,1.90  \\
                                              &
            (60.0) &  (40.5) &
            (36.2) &  (37.0) &
            (97.9) &  (27.5) &
            (75.2) &  (50.1) \\
F$_{X,1}$ ($10^{-13}$ erg cm$^{-2}$ s$^{-1}$)  &
           \,\,0.55  & \,\,0.47  &
           \,\,0.58  & \,\,0.49  &
           \,\,0.55  & \,\,0.50  &
           \,\,0.54  & \,\,0.48  \\
                                              &
            (0.79) &  (0.88) &
            (0.87) &  (0.91) &
            (1.01) &  (1.06) &
            (1.19) &  (1.40) \\
$\log L_X$ (erg s$^{-1}$) &
           34.64 & 34.47 & 34.42 & 34.43 &
           34.85 & 34.30 & 34.74 & 34.56 \\
$\log L_{X.1}$ (erg s$^{-1}$) &
           32.76 & 32.80 & 32.80 & 32.82 &
           32.86 & 32.89 & 32.94 & 33.00 \\
\hline

\end{tabular}

{\it Note}.
Results from  fits to the EPIC
spectra of \WRE. The labels MOS and MOS$+$pn denote which spectra 
were fitted.  The \xspec models are: 
$wabs*apec+wabs*apec$ (Model A); 
$wabs*(phabs*apec+phabs*apec)$ (Model B);
$wabs*powerlaw +$ $wabs*powerlaw$ (Model C); 
$wabs*(phabs*powerlaw+phabs*powerlaw)$ (Model D).
Note that both model components have the same spectral shape (see
text for details).
Tabulated quantities are the neutral hydrogen absorption column
density (N$_{H}$), helium absorption column density (N$_{He}$) for
the `wind' components, plasma temperature (kT),
emission measure ($\mbox{EM} = \int n_e n_{He} dV $),
photon power-law index ($\Gamma_{pow}$), power-law model
normalization (norm$_{pow}$), 
the observed X-ray flux (F$_X$) in the
0.5 - 10 keV range followed in parentheses by the unabsorbed value
and the X-ray luminosity ($L_X$).
F$_{X,1}$ and $L_{X.1}$ denote the values for the first 
model component, peaking at lower energies.
The abudnaces for the thermal models are those typical for
the WO stars 
(by number: 
H~$=0.00$, He~$=0.266$, C~$=0.213$, N~$=0.00$, O~$=0.497$,
Ne~$=9.69\times10^{-3}$, Mg~$=1.26\times10^{-2}$,
Si~$=6.73\times10^{-4}$, S~$=1.47\times10^{-4}$,
Fe~$=3.69\times10^{-4}$; \citealt{vdh_86}).
The values for the emission measure and the X-ray luminosity are for a
reference distance of d~$=7.77$~ kpc.
Errors are the $1\sigma$ values from the fits.

\end{table*}

\begin{figure*}
\centering\includegraphics[width=1.89in,height=2.835in,angle=-90] {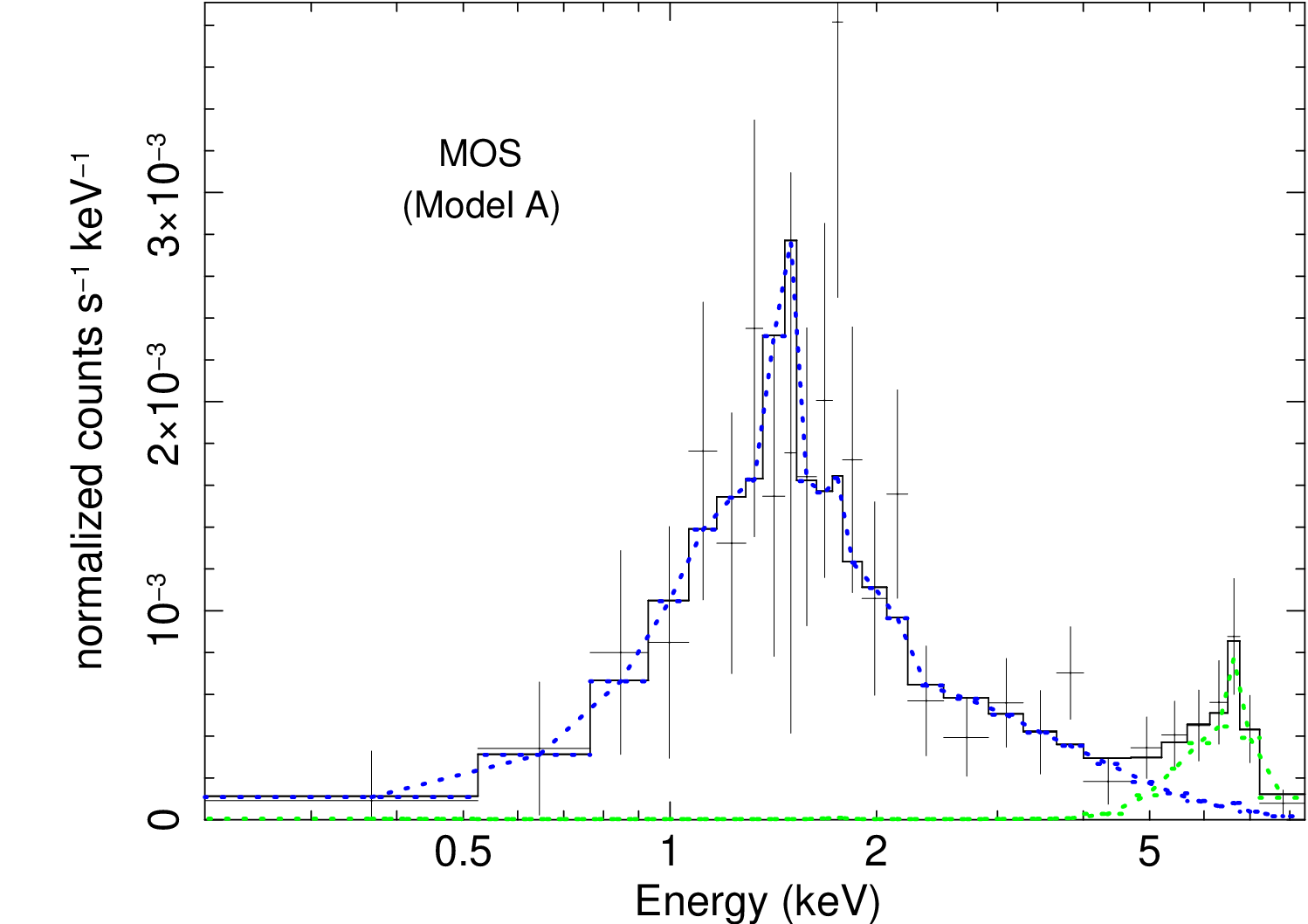}
\centering\includegraphics[width=1.89in,height=2.835in,angle=-90] {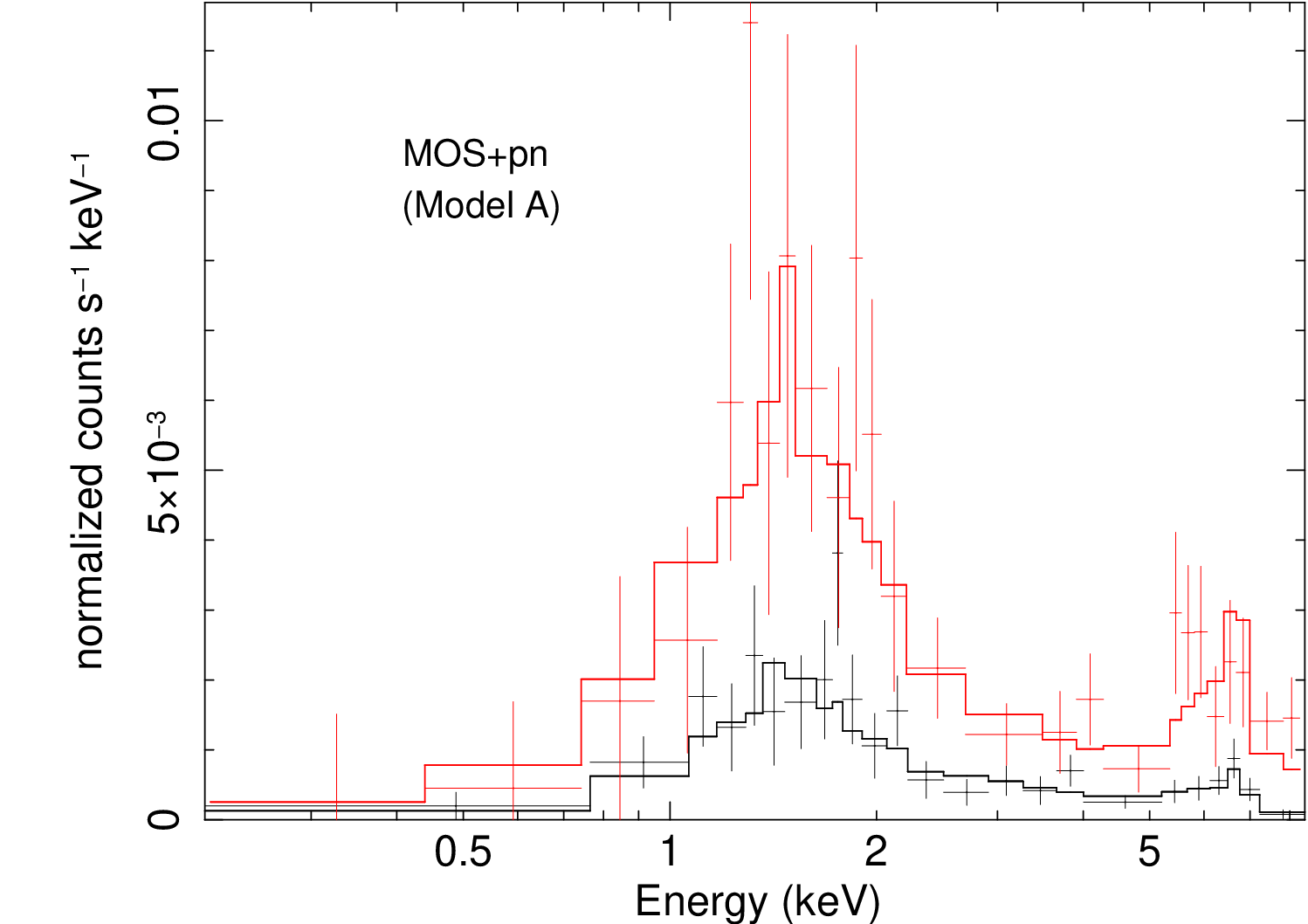}
\centering\includegraphics[width=1.89in,height=2.835in,angle=-90] {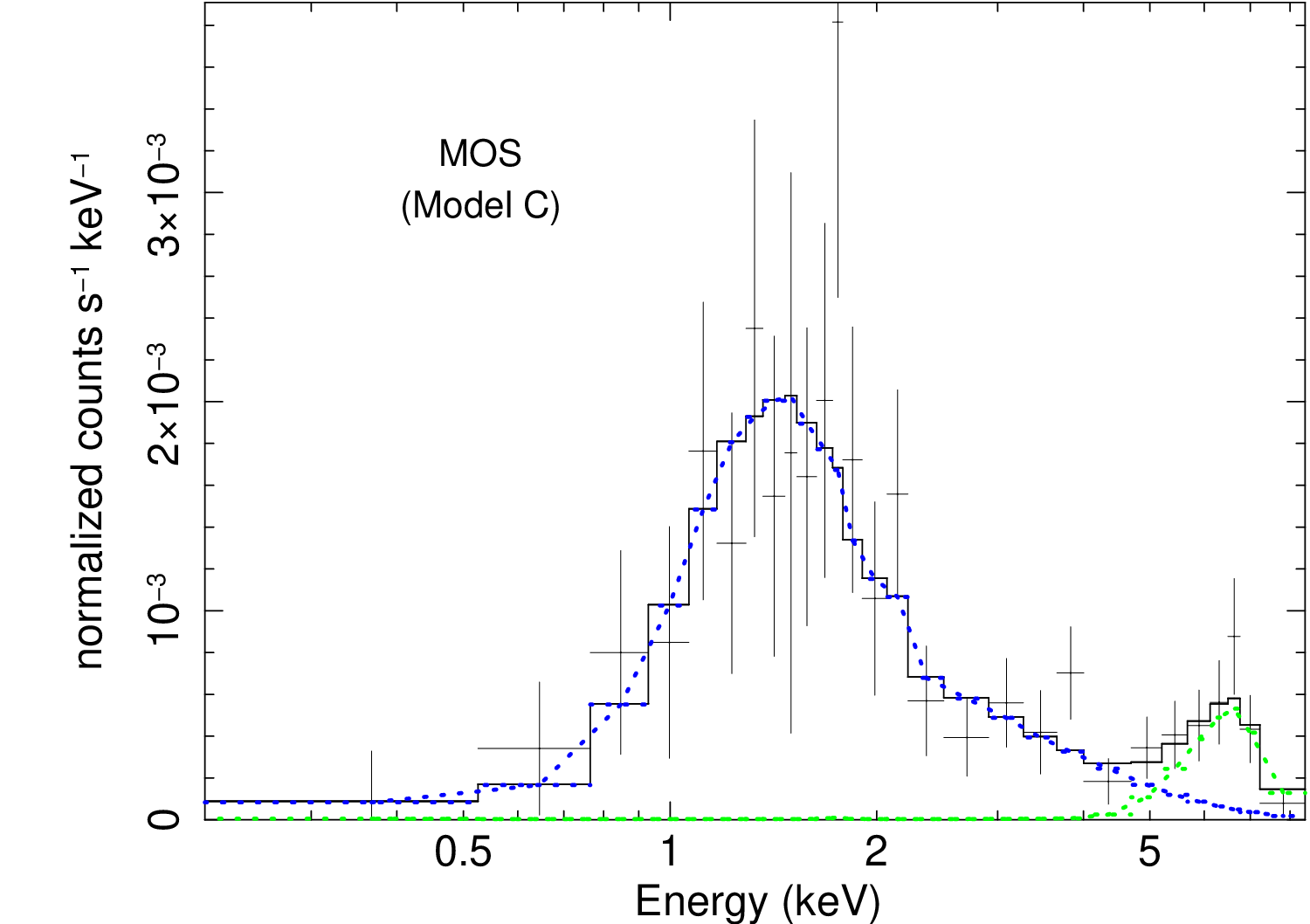}
\centering\includegraphics[width=1.89in,height=2.835in,angle=-90] {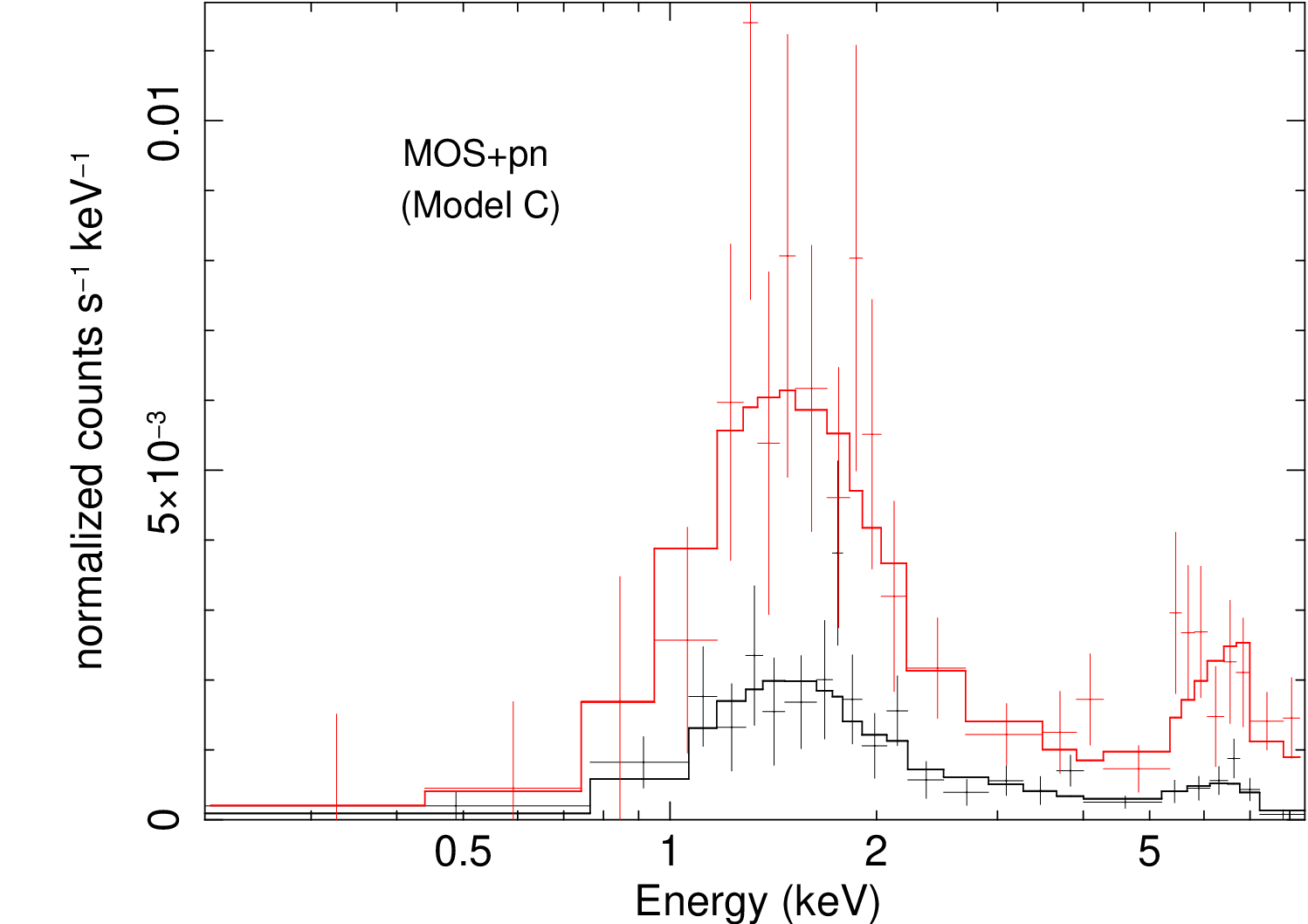}
\caption{
The background-subtracted EPIC spectra of \WR overlaid with some
best-fitting models (Table~\ref{tab:fits}). In the case of 
simultaneous fit to the MOS and pn spectra, the latter is plotted in 
red. For presentation purposes, the spectra were slightly re-binned 
with respect to the original binning of 20 counts per bin used in the 
fits. The model components are shown in blue and green colour.
}
\label{fig:spectra}
\end{figure*}

\section{Results}
\label{sec:results}
As apparent from  Fig.~\ref{fig:spectra}, the X-ray spectra are nearly
featureless and are characterized by two broad peaks or `bumps',
one at lower energies (peaking between 1 and 2 keV) and another  at 
higher energies (maximum emission between 5 and 7 keV). Other than a 
possible contribution from the Fe K complex near 6.67 keV, there are 
no bright spectral lines that would clearly indicate thermal emission. 
We  thus explored both thermal (optically-thin plasma) and
non-thermal (power-law) models in \xspecE.

For the thermal models, we adopted typical WO star abundances  from 
\citet{vdh_86}.
The emission plasma components were subject to common X-ray absorption
(model {\it wabs} in \xspecE) representative of the interstellar
medium (ISM). Some additional absorption, as expected from the winds,
was also required to acceptably fit the total spectrum. 
Typical WO star
abundances were adopted for the wind absorption which was modeled 
using {\it phabs} model in \xspecE. Due to the rather low number of 
X-ray counts and absence of strong lines, all the abundances were 
kept fixed at their adopted values during fitting.

The one-component model subject only to ISM X-ray absorption did not
provide a good fit to the observed MOS spectrum of \WRE. The same was 
true for the two-component model when using the same ISM absorption for
each component.

The two-component model with individual X-ray absorptions for each 
emission component gave a very good fit to the MOS spectrum.
However, the best-fit plasma-temperature (or power-law index) values 
of the different emission components overlapped at the  $1\sigma$ 
level. We thus adopted a common value for these parameters in 
the final model fits, which means that the spectral shape of the
emission components was the same but the two components differed in
their respective  amounts of thermal plasma (or nonthermal emission) 
and in their X-ray absorptions.
When exploring the cases with wind absorption, we kept the common 
ISM X-ray absorption fixed to the values that correspond to the
optical extinction of \WR (Section \ref{sec:wrstar}).
As an additional check of our global modelling, we fitted the MOS and
pn spectra of \WR simultaneously by making use of the same models  
described above.

Table~\ref{tab:fits} and Fig.~\ref{fig:spectra} present the
corresponding results from the fits to the \XMM spectra of \WRE.
The fit results can be briefly summarized as follows.
Both thermal and non-thermal models are successful in reproducing the
X-ray spectrum.
In the case of thermal models, a relatively hot plasma (kT $> 2$~keV) 
is required to be present in the X-ray emitting region.
The first emission component in the models is subject to X-ray 
absorption that is in general consistent with the optical extinction
toward \WRE , while the second emission component is more heavily
absorbed. If we adopt a common ISM absorption for both emission
components, then only the  second component requires some 
additional wind absorption (see Models B and D in
Table~\ref{tab:fits}). For illustration of the results from the model
fits with a common ISM absorption, we show only the case with the 
upper limit for the range of values that correspond to the optical 
extinction (Section \ref{sec:wrstar}). We note that if the lower
limit for the ISM absorption was used, then the derived model 
parameters were within their respective $1\sigma$ confidence 
intervals.

The X-ray luminosity of \WR, $L_X > 10^{34}$ erg s$^{-1}$, 
is quite high when compared to other WR stars in the Galaxy.
The more heavily absorbed component responsible for
the observed emission in the 5 - 7 keV range dominates the 
total luminosity in the 0.5-10 keV energy band.

\section{Discussion}
\label{sec:discuss}
Some of the results from the X-ray spectral fits may be affected
by the fact that there are relatively few counts in the \XMM spectra
(Section~\ref{sec:observations}). Nevertheless, the following 
physical picture emerges. There is likely  some distribution of 
thermal X-ray plasma (or non-thermal emission) with similar plasma 
characteristics and this X-ray emission region is only partially 
absorbed. Alternatively, there may be  two different X-ray sources 
with about the same plasma characteristics but one of them is more 
heavily absorbed. If the absorption is due to the stellar wind, that 
source must be located closer to the WO star while the other 
less-absorbed source is further from the star and is not subject 
to significant wind absorption.

If the X-rays from \WR are from CSW shocks in the WR$+$O binary, 
we expect a temperature-stratified interaction region. Thus, the 
derived values of the plasma temperature in our fits represent a mean 
temperature for the shocked plasma. 
For the case of typical WO abundances, the postshock plasma
temperature is kT~$ = 3.36 ~V_{1000}^2$~keV, where $V_{1000}$ is
the shock velocity in units of 1000\kms.
Given the stellar wind velocity of 
4500\kms (Section~\ref{sec:wrstar}), high temperatures at  kT $> 2$ keV 
are realistic for shocks in WO plasmas.  Very high shock temperatures  
could be achieved even if the
winds have not reached their terminal velocities before colliding,
which could be the case in this close WO$+$O binary 
(orbital period of 4.619 d; Section~\ref{sec:wrstar}).

Even though the  high plasma temperatures (kT $> 2$ keV) determined 
from spectral fits could be taken as evidence for colliding stellar 
winds, it is worth keeping in mind that such high temperatures have 
also been detected in presumably single WN stars 
(\citealt{sk_10}, 2012) and in very close binaries like 
WR 46 (see \citealt{gosset_11}; \citealt{zh_12}) and 
WR 155 (CQ Cep; see \citealt{sk_15}). So, CSWs  may not be the only  
mechanism  producing very hot plasma in WR stars.

However, as already noted, the spectral fits of \WR cannot clearly 
distinguish between thermal and nonthermal models. This ambiguity is 
due in part to the low number of spectral counts and higher 
signal-to-noise spectra capable of determining whether weak emission 
lines are present will be needed to discriminate between thermal and 
nonthermal emission. A similar ambiguity was found for the presumably 
single WO star WR 142 whose X-ray emission is even fainter and shows 
no strong lines \citep{sokal_10}. \WR  and WR 142 also reveal other 
similarities, e.g. both have a hard heavily-absorbed spectral component.
A notable difference is that the X-ray luminosity of WR 142 is less 
than 1\% of that of \WRE. The binary nature of \WR may be largely 
responsible for this, but distance uncertainties
could also contribute.

The X-ray luminosity of \WR ($L_X > 10^{34}$ 
erg s$^{-1}$, Table~\ref{tab:fits}) may be the highest 
amongst the close WR$+$O binaries (orbital periods less than 
$\sim20$ d) in the Galaxy studied so far (e.g., see 
\citealt{zh_12}; \citealt{sk_15}; \citealt{naze_08}), provided the 
adopted distance of 7.77 kpc is not significantly overestimated. 
The high X-ray luminosity must be accommodated by any plausible 
theoretical emission models.

We note that other close WR binaries with very
high X-ray luminosities may exist in the Galaxy.
\citet{gagne_12} have compiled a list of possi-
ble detections of colliding wind binaries with \Chandra, \XMM
and \RosatE. High X-ray luminosity candidates are WR28, WR29, 
WR43a, WR43c and WR101k. However, caution is needed when considering 
the available data on these objects. For example, WR43a and WR43c are 
located in the core of the Galactic starburst region NGC 3603 and
thus cannot be spatially resolved in X-rays (that is their X-ray 
spectra are contaminated by the near-by sources). Similarly, WR101k 
cannot be resolved since it is in the centre of the Galaxy only 
1.8$''$ from Sgr A (SIMBAD). The data on WR28 and WR29 are from \Rosat 
observations and only upper limits can be derived on their observed 
fluxes. Nevertheless, these close WR binaries which may have
extremely high X-ray luminosities deserve more attention and 
in-depth X-ray studies.

Finally, a distinguishing feature of \WR is its unusual spectral shape 
with two broad separated peaks. After examining the X-ray spectra of 
other WR stars we were able to identify another object with a 
similarly shaped spectrum, the close WR$+$O
binary WR 79.

\subsection{Comparison with WR 79}
\label{subsec:comparison}
\begin{figure}
\centering\includegraphics[width=2.14in,height=3in,angle=-90] {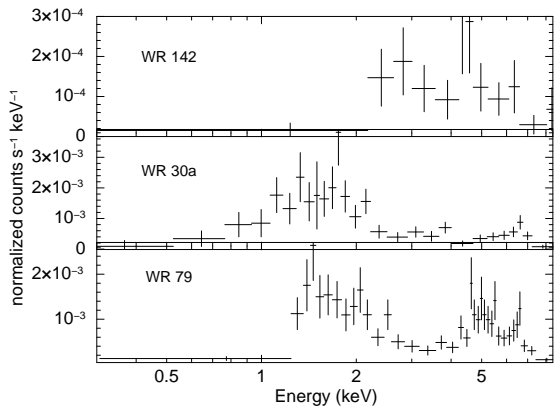}
\caption{
The background-subtracted  spectra of \WR (\XMM EPIC-MOS),
WR 79 (\Chandra ACIS-S, ObsId 5372, see \citealt{zh_12})
and
WR 142 (\Chandra ACIS-I, ObsId 9914; based on our own spectral
extraction using version 4.7 of Chandra Interactive Analysis of
Observations, CIAO; http://cxc.harvard.edu/ciao/).
}
\label{fig:siblings}
\end{figure}

WR 79 is a WC$+$O binary with an orbital period of 8.28 d at a 
distance of 1.99 kpc \citep{vdh_01}. Figure~\ref{fig:siblings} 
compares its \Chandra  X-ray spectrum with  that of \WRE ~and WR 142.
We note that WR 79 was also observed by \XMM but its spectrum
is contaminated by two nearby sources within 7$''$ - 8$''$
that are seen in the higher spatial resolution {\em Chandra}
image (for details,  see Appendix A in \citealt{zh_12}).

Two broad peaks are present in both WR 79 and  \WRE, but only the 
higher energy peak is present in the apparently single WO star WR 142.
No prominent spectral lines are seen with the possible exception of
Fe emission in the 6-7 keV range in WR 79 and a hint of the same
in WR 30a. But faint emission lines could be missed since the number 
of counts in all three spectra is quite limited, with only $\sim 380$ 
source counts in WR 79 (\citealt{zh_12}) and  $\sim$46 counts in 
WR 142 \citep{sokal_10}.

The X-ray spectrum of WR 79  was well matched by a
two-component optically-thin plasma model with individual wind
absorptions (see Table 3 and Fig. 1 in \citealt{zh_12}). As the global
fits showed, the second model component due to hot plasma with kT $> 2$ keV
is subject to much higher wind absorption, which is also the case
with \WRE. And the more absorbed component, whose emission produces
the broad peak between 4 and 7 keV in the spectrum, dominates the total
luminosity in the 0.5-10 keV energy range. We did this estimate by
making use of exactly the same spectral data (no re-processing)
and models as in \citet{zh_12}. Despite these similarities, there is an
interesting difference: the X-ray luminosity of WR 79 is less than \WR 
by approximately one order of magnitude.

\section{Conclusions}
\label{sec:conclusions}
In this work, we presented and analyzed archival  \XMM data of \WR 
which provide the first X-ray spectra of this WR$+$O binary that 
harbours a rare  oxygen-rich Wolf-Rayet star. The main results and 
conclusions are as follows.

(i) The X-ray spectrum of \WR shows no strong spectral lines, although
some Fe emission between 6 - 7 keV may be present and other weak lines 
may have been missed due to the low number of source counts in the 
spectrum. The spectrum can be acceptably fitted by either thermal or 
nonthermal emission models. 

(ii) The most distinguishing feature in the \WR spectrum is the 
presence of two broad but well-separated peaks,  one peaking at energies 
between 1 and 2 keV and the other between 5 and 7 keV. A two-component
model is required to match the observed spectrum. The spectral component 
seen in the higher energy peak  is considerably more absorbed and 
dominates the intrinsic X-ray luminosity in the 0.5-10 keV range.

(iii) For the currently accepted distance of 7.77 kpc, the X-ray
luminosity of \WR is $L_X > 10^{34}$ erg s$^{-1}$, making it the 
most X-ray luminous object amongst the close WR$+$O binaries 
(orbital periods less than $\sim20$ d) in the Galaxy studied 
so far.

(iv) On the basis of the double-peaked spectral shape  and
other X-ray characteristics, we have identified  two WR binaries
that are quite similar: \WR and WR 79 (WC$+$O). 

(v) Future observations of both single and binary WO stars are needed
to obtain higher signal-to-noise X-ray spectra capable of determining 
if faint emission lines are present and thereby  distinguishing 
between thermal and nonthermal emission processes.

\section{Acknowledgements}
This research has made use of data and/or software provided by the
High Energy Astrophysics Science Archive Research Center (HEASARC),
which is a service of the Astrophysics Science Division at NASA/GSFC
and the High Energy Astrophysics Division of the Smithsonian
Astrophysical Observatory. 
This research has made use of the NASA's Astrophysics Data System, and
the SIMBAD astronomical data base, operated by CDS at Strasbourg,
France.
This work is based on observations obtained with XMM-Newton, an ESA 
science mission with instruments and contributions directly funded 
by ESA Member States and the USA (NASA).

{}

\bsp

\label{lastpage}

\end{document}